# Chip-scale Spontaneous Quasi-Phase-Matched Micro-Racetrack Resonator

Tingge Yuan[1, ‡], Yi'an Liu[1, ‡], Xiongshuo Yan[1], Haowei Jiang[1], Hao Li[1], Rui Ge[1], Yuping Chen[1, *], Xianfeng Chen[1,2,3*] and Qiang Lin[4]

[1] *Institute of Optical Science and Technology, School of Physics and Astronomy, School of Physics and Astronomy, Shanghai Jiao Tong University, 800 Dongchuan Road, Shanghai 200240, China*

[2]*Shanghai Research Centre for Quantum Sciences, Shanghai 201315, China*

[3]*Collaborative Innovation Centre of Light Manipulations and Applications, Shandong Normal University, Jinan 250358, China*

[4]*Department of Electrical and Computer Engineering, University of Rochester, Rochester, New York 14627, USA*

*Email: ypchen@sjtu.edu.cn*

*Email: xfchen@sjtu.edu.cn*

‡*These authors contributed equally to this work.*

**Abstract**

Due to their capacity for non-classical light generation, high-efficiency second-order nonlinear parametric processes play an important role in quantum photonic technology, and chip-scale realization of these processes is recognized as the key to building efficient light sources for integrated quantum photonic circuits. To achieve ultra-high nonlinear conversion efficiency, traditional method uses quasi-phase matching (QPM) technology. However, QPM requires electric field poling, which is incompatible with the CMOS fabrication process, and this hinders the wafer-scale production of integrated photonic circuits. In this paper, we demonstrate efficient spontaneous quasi-phase matched (SQPM) frequency conversion in a micro-racetrack resonator. Our approach does not involve poling, but exploits the anisotropy of the ferroelectric crystals to allow the phase-matching condition to be fulfilled spontaneously as the TE-polarized light circulates in a specifically designed racetrack resonator. SQPM second harmonic generation is observed with a normalized intracavity conversion efficiency of 0.85%/$W$, corresponding to the 111$st$-order QPM. This could theoretically reach 186,000%/$W$ by first-order QPM. In this case such high intracavity conversion efficiency can be implemented in practice with an optimized outward coupling. Our configurable SQPM approach will benefit the application of nonlinear frequency conversion in chip-scale integrated photonics with CMOS-compatible fabrication processes, and is applicable to other on-chip nonlinear processes such as quantum frequency conversion or frequency-comb generation.

# Introduction

As the most fundamental nonlinear optical phenomena, second-order nonlinear parametric processes have attracted a great deal of interest since the laser was invented in the 1960s. They have been developed into a powerful tool for generating various non-classical light sources, such as entangled photon pairs[1-3] and squeezed light[4,5] in quantum photonic networks, and more generally to provide a flexible frequency conversion to transfer the quantum state of light between different wavelengths[6,7]. The phase-matching condition is necessary to achieve a high-efficiency nonlinear interaction[8]; however, perfect fulfilment of this normally requires strict conditions on both the dispersion property of the nonlinear medium and the orientation of the crystal axis. Quasi-phase matching (QPM) is another approach to achieving effective nonlinear interactions[9-11] beyond birefringent phase matching[8]. This requires the sign or magnitude of the second-order nonlinear susceptibility to be periodically modulated, thus the phase mismatch between the interacting waves can be intermittently compensated and the intensity of the newly generated frequency component will grow continuously. To date, the most common and mature approach to constructing a QPM structure is electric-field-induced ferroelectric domain inversion, (i.e., electric field poling[12]), via which the sign of the nonlinear susceptibility can be totally inverted.

In recent years, various chip-integrated photonic elements have been reported[13-25]; of these, high-quality micro-resonator has emerged as a promising platform of optical nonlinear interactions due to the significant enhancement of the light-material interactions[26-29]. With the assistance of QPM, ultra-high conversion efficiency has been achieved in the micro-ring resonators[30-33], which is expected to be an ideal substrate for high-performance quantum light sources in integrated quantum photonic circuits. However, in view of the available fabrication processes, the inevitable step of electric field poling for QPM still remains a challenge in terms of integrating QPM-assisted devices with other components on integrated quantum photonic circuits through a standard CMOS-compatible process. To fully explore the potential of wafer-scale quantum photonic chips, there is an urgent need to find a novel CMOS-compatible approach to realize the high-efficiency on-chip nonlinear frequency conversion processes.

In this paper, we propose and demonstrate spontaneous quasi-phase matching (SQPM), a poling-free mechanism for achieving phase-matched high-efficiency frequency conversion in a micro-racetrack resonator on X-cut thin-film lithium niobate. By dispersion engineering and the design of the crystal orientation, a periodically inversed domain appears spontaneously as the TE- polarized light circulates in the resonator, in a similar way to the periodically poled QPM structure, although no more poling steps are required. As a result, the realization of SQPM can be CMOS-compatible. In experiments, an enhanced second harmonic generation (SHG) with a normalized intracavity conversion efficiency of 0.85%/$W$ is observed. Theoretical predictions indicate that this value can be further improved to about 186,000%/$W$ by optimizing the racetrack structure and improving the quality (Q) factor of the resonator. The SHG bandwidth and the fabrication tolerance are also discussed. In addition to the example presented in this work, SQPM can be employed for other second-order nonlinear frequency conversion processes, such as spontaneous parametric down-conversions, optical parametric oscillations, and even the generation of micro-frequency combs, which will promote the application of integrated nonlinear photonics in both the classical and quantum regimes.

# Results

**Spontaneous Quasi-Phase-Matching in Micro-racetrack Resonator**

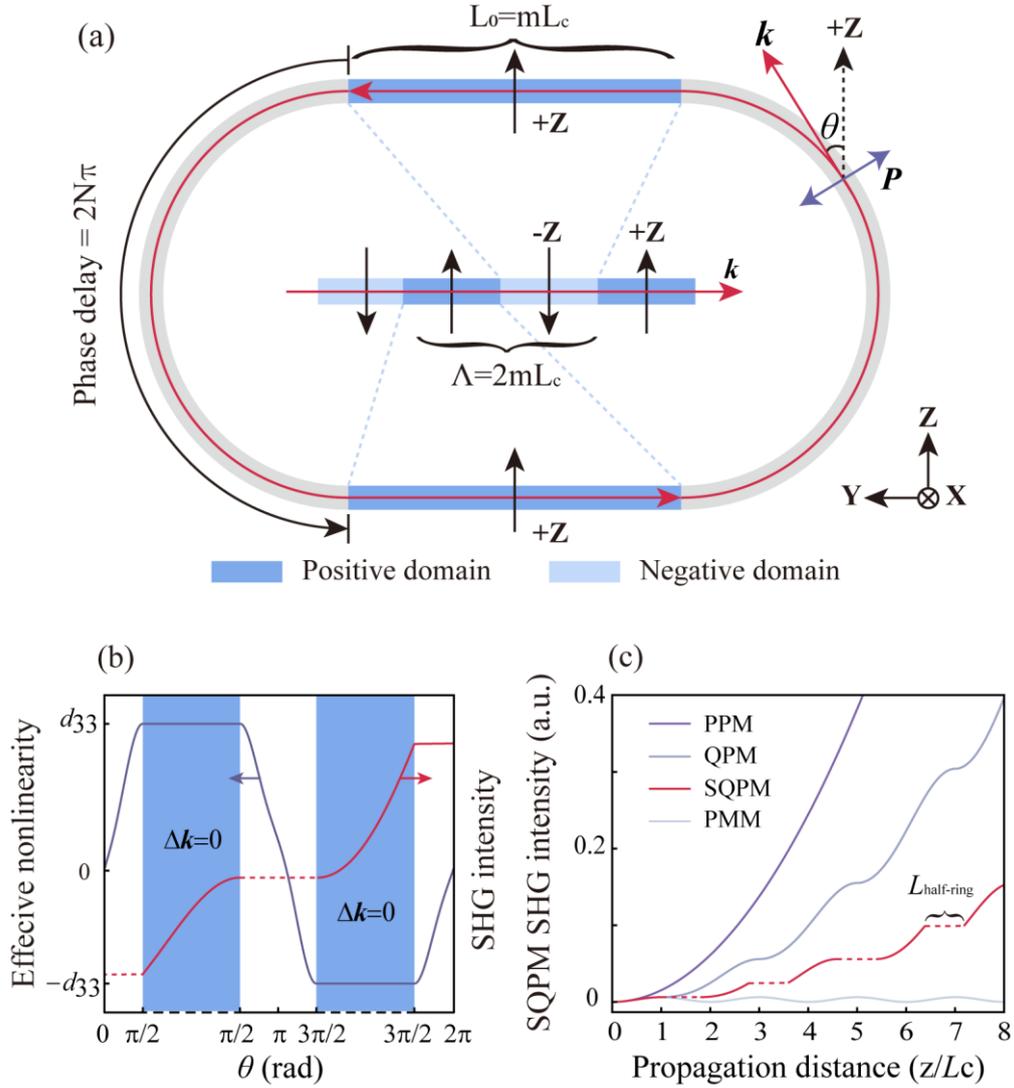

**Fig 1. Principle of the spontaneous quasi-phase matching.** (a) Schematic of the SQPM in a poling-free racetrack resonator on X-cut thin film lithium niobate. Inset: traditional configuration of QPM by electric field periodic poling. (b) Relationships between the effective nonlinear coefficient and SQPM SHG intensity with the azimuth angle $\theta$, varying from 0 to $2\pi$. (c) SHG intensity with the propagation distance under different phase-matching conditions. PPM: perfect phase-matching, PMM: phase mismatching, $L_{\text{half\_ring}}$ is the perimeter of the half-ring waveguide. The Y-axis has been normalized to the maximum intensity obtained at $8L_c$ under PPM.

Lithium niobate is an outstanding optical material due to its huge second-order nonlinear susceptibility and the capability of being poled[34]. Furthermore, with the assistance of QPM technology, periodically poled lithium niobate can be used to realize the ultra-high efficiency of the nonlinear frequency conversion process. As a result, we choose the lithium niobate based SQPM as a typical example, and show how we can achieve a similar result of QPM SHG in a poling-free way.

As Fig. 1(a) showed, under the traditional configuration of QPM based on the electric field poling, the domain orientation (i.e., the Z-axis of lithium niobate) is periodically inverted along the direction of propagation of the light with a period of Λ. For the phase-matching condition, Λ should satisfy the relation:

$$\Lambda = \frac{2\pi m}{\Delta k} = 2mL_c, \quad (1)$$

where $m$ is a positive odd, representing the order of the QPM, $\Delta k$ is the phase mismatch in the nonlinear interaction. For a typical SHG process, $\Delta k = k_{SH} - 2k_{FW}$, where $k_{SH}$ and $k_{FW}$ is the wave vector of the second harmonic (SH) wave and fundamental wave (FW), respectively. $L_c$ is the coherent length of the nonlinear interaction.

This traditional mechanism for QPM can be summarized as a periodically inverted domain and an unchanged direction of propagation for the light. In view of the dependence of the effective nonlinear coefficient on the relative angular relation between the wave vector and the crystal axis, it is possible to obtain a result equivalent to the periodically poled lithium niobate with the domain orientation unchanged if the propagation direction of the light is periodically reversed. This configuration can be perfectly realized in a micro-racetrack resonator on X-cut lithium niobate, where the effective nonlinear coefficient is given by

$$d_{eff} = -d_{22} \cos^3 \theta + 3d_{31} \cos^2 \theta \sin \theta + d_{33} \sin^3 \theta. \quad (2)$$

Here $\theta$ is defined as the azimuthal angle between the wave vector $\mathbf{k}$ and the Z-axis. Each time the TE-polarized light travels through one of the straight waveguides, the direction of propagation rotates through 180°, corresponding to inversion of the sign of the effective nonlinear coefficient. Thus, we can draw an analogy between the straight waveguides in micro-racetrack resonator to the adjacent domains in a period of the periodically poled lithium niobate, and design a straight waveguide length $L_0$ following the principle of QPM. In addition, the half-ring section that connects the straight waveguides needs a special design to maintain the phase relation between the FW and SH, meaning that the phase delays of both the FW and SH in the half-ring waveguide should be equal to integer multiples of $2\pi$.

Based on the Eq. (2), the variation in the effective nonlinear coefficient $d_{\text{eff}}$ and the SHG intensity with $\theta$ within an SQPM period is shown in Fig. 1(b). We can see that $d_{\text{eff}}$ undergoes a dramatic oscillation in the phase-mismatched white regions, while the SHG intensity increases continuously in the blue regions where the QPM condition is fulfilled. A comparison with different phase-matching conditions is shown in Fig. 1(c). It can be seen from the red curve in Fig. 1(c) that due to the existence of the half-ring waveguide, the rise in the SHG intensity under the SQPM condition suffers from periodic stagnation over a length of $L_{\text{half\_ring}}$, which will ultimately influence the SHG conversion efficiency.

**Design of the SQPM micro-racetrack resonator**

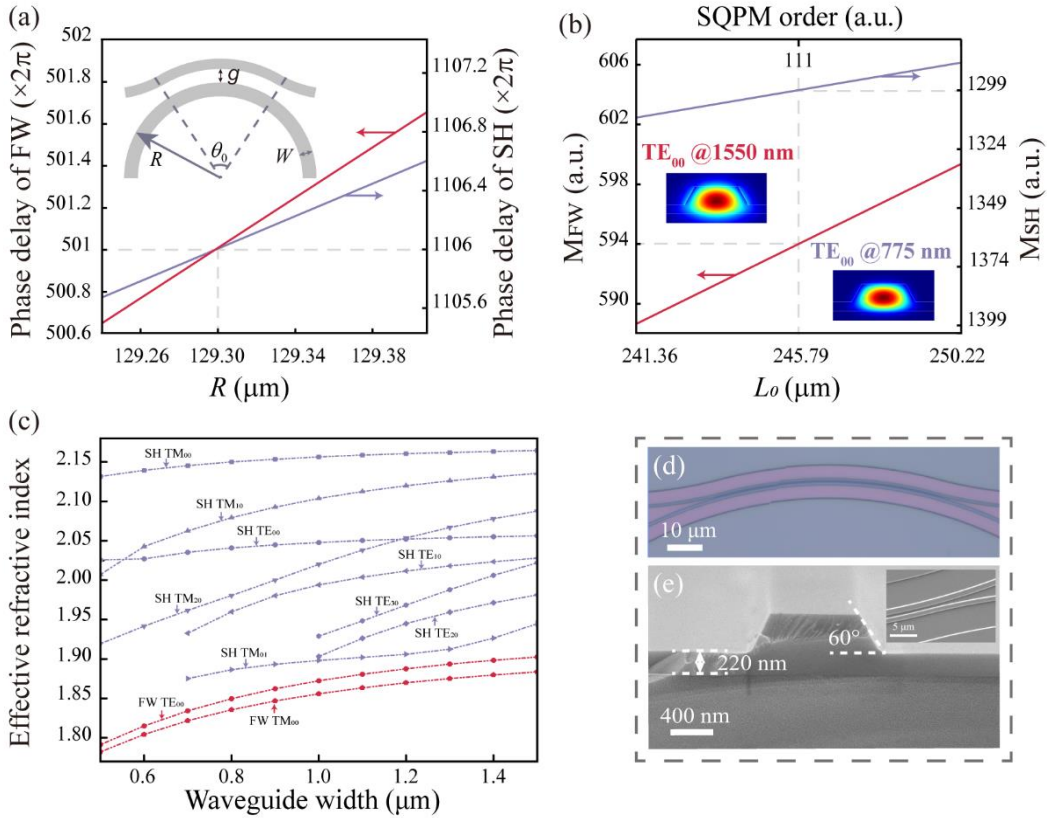

**Fig. 2 Design of the SQPM micro-racetrack resonator.** (a) Phase delays in a half-ring waveguide with the outside radius $R$ for FW and SH. Insert: diagram of the pulley coupling waveguide. (b) Wavenumbers of the straight-section of the SQPM micro-racetrack resonator with the straight waveguide length $L_0$ for FW and SH. Insert: simulated filed distribution of $TE_{00}$ mode at FW and SH, respectively. (c) Effective refractive indices of the FW and SH modes supported by the micro-racetrack resonator. The order number of each mode is determined by the mode distribution in the waveguide with a width of 1 $\mu m$. (d) Optical microscopy image of the pulley coupling waveguide. (e) Cross section and coupling region of the micro-racetrack resonator under a scanning electron microscope.

As a typical example, consider a micro-racetrack resonator for SHG with a FW wavelength of 1550 nm, in which both the FW and SH are in the TE fundamental mode for the largest mode overlap. We started the design with the half-ring waveguides. The cross section of the waveguide was defined with a top width of 1 μm, a thickness of 0.38 μm, and a side-wall angle of 60° for the single-mode condition; based on these values, we calculated the phase delays for the fundamental TE modes (TE$_{00}$) at the FW and SH wavelengths in the half-ring waveguide with varying outside radius $R$, as shown in Fig. 2(a). For $R = 129.30$ μm, both the phase delays are equal to integral multiples of $2\pi$. In the next, the straight waveguide length $L_0$ is calculated following the QPM principle described above, which requires

$$L_0 = mL_c = \frac{m\lambda}{4(n_{\text{eff,SH}} - n_{\text{eff,FW}})}, \quad (3)$$

where $n_{\text{eff,FW(SH)}}$ is the effective refractive index of the fundamental mode of FW (SH) in the straight waveguide. In addition, the resonance of FW and SH in the micro-racetrack resonator requires

$$\begin{cases} 2L_0 n_{eff,FW} = M_{FW}\lambda \\ 2L_0 n_{eff,SH} = M_{SH}\lambda/2 \end{cases} \quad (4)$$

where M$_{FW}$ indicates the azimuthal mode number in the straight waveguide for FW. In Fig. 2(b), this calculation gives $L_0 = 245.79$ μm, corresponding to the 111st-order QPM. Finally, we use a pulley coupling waveguide[35] and a pair of grating couplers to connect the micro-racetrack resonator with the laser source. To maximize the coupling efficiency of the TE$_{00}$ mode for the FW band, the gap $g$ and the width of the coupling waveguide $W_{\text{wg}}$ were designed to be 0.8 μm and 0.61 μm, respectively. The central angle $\theta_0$ was 30°. Fig. 2(d) and (e) show the optical and scanning electron microscopy images of the fabricated micro-racetrack resonator based on our design, where the lithium niobate film is etched to a total depth of 380 nm, leaving a slab of 220 nm.

Here, to ensure that only SQPM enabled SHG will take place in our designed micro-racetrack resonator, effective refractive indices for the possible modes at FW and SH are calculated in Fig. 2(c) versus the top width of the waveguide. Here only fundamental modes can be supported by our waveguide at FW band, while high-order modes can exist for the SH band. Obviously, no intersection exists between the different modes of FW and SH, which means the modal phase matching cannot occur in our designed waveguide.

At last, based on the coupled-mode theory[36] and the generic description of SHG in a whispering gallery resonator[37], we derive an expression for the normalized intracavity conversion efficiency of SHG in an SQPM micro-racetrack resonator as follows (see Supplemental Material):

$$\eta = \eta_0 \frac{\iint |A_{FW}|^4 d\sigma_{SH}}{(\iint |A_{FW}|^2 d\sigma_{FW})^2}, \quad (5)$$

where $\eta_0 =$

$$\frac{128 Q_{l,SH}^2 d_{33}^2 L_c^2}{(m_{\text{azi,SH}} + M_{SH})^2 \pi^2 \lambda^2 c\varepsilon_0 n_{\text{eff,FW}}^2 n_{\text{eff,SH}}}, \quad (6)$$

$A_{FW}$ is the intracavity amplitude of FW, and $\sigma$ is the cross-sectional area of the fundamental mode in the straight waveguide ($\sigma_{FW}$ for FW, and $\sigma_{SH}$ for SH). $Q_{l,SH}$ is the loaded Q factor of the micro-racetrack resonator in the SH band. $\lambda$ is the vacuum wavelength of FW. $m_{azi,SH}$ and $M_{SH}$ are the azimuthal mode numbers of SH in the half-ring and straight section of the racetrack resonator, respectively. $n_{eff,FW(SH)}$ is the effective refractive index of the fundamental mode of FW (SH) in the straight waveguide. It should be noticed that the main reason we study the intracavity instead of the on-chip conversion efficiency here is to exclude the influence from the coupling and focus on the behaviour of SQPM enabled SHG itself. From Eq. (6), the intracavity conversion efficiency directly dependent on $Q_{l,SH}$ and the size of the micro-racetrack resonator, which is represented by $m_{azi,SH}+M_{SH}$ here. We will give a profound discussion on these factors in the next sections.

**Spontaneous Quasi Phase Matched Second Harmonic Generation**

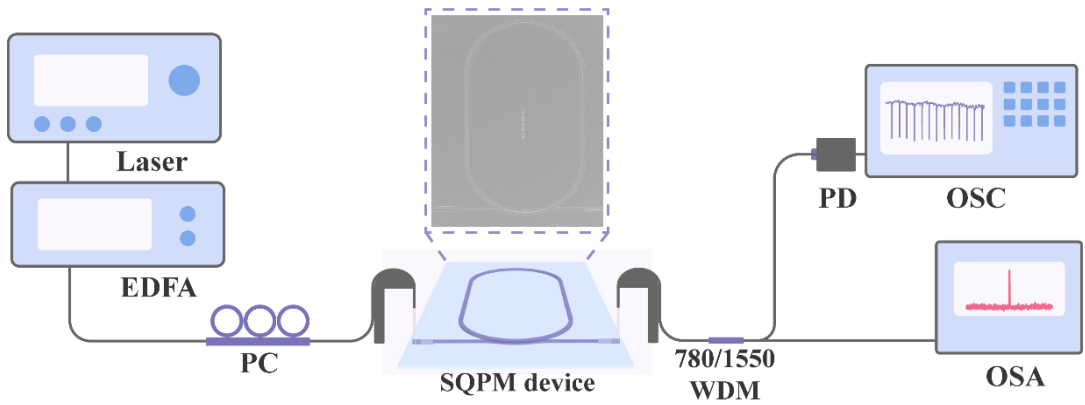

**Fig. 3 Experimental setup.** EDFA: erbium-doped optical fibre amplifier, PC: polarization controller, WDM: wavelength division multiplexing, PD: photodetector, OSC: oscilloscope, OSA: optical spectrum analyser.

In this section, SQPM SHG is experimentally demonstrated. The experimental setup is shown in Fig. 3. An infrared tunable laser (New Focus TLB-6728) served as the laser source of FW. Before being coupled to the lithium niobate film, the input laser was amplified by an erbium-doped optical fibre amplifier, and then adjusted by a polarization controller. At the output port, the FW and SH were separated by a wavelength division multiplexing. A photodetector connected to an oscilloscope was used to measure the transmission spectrum of the FW, while an optical spectrum analyser was used to detect the SH signal.

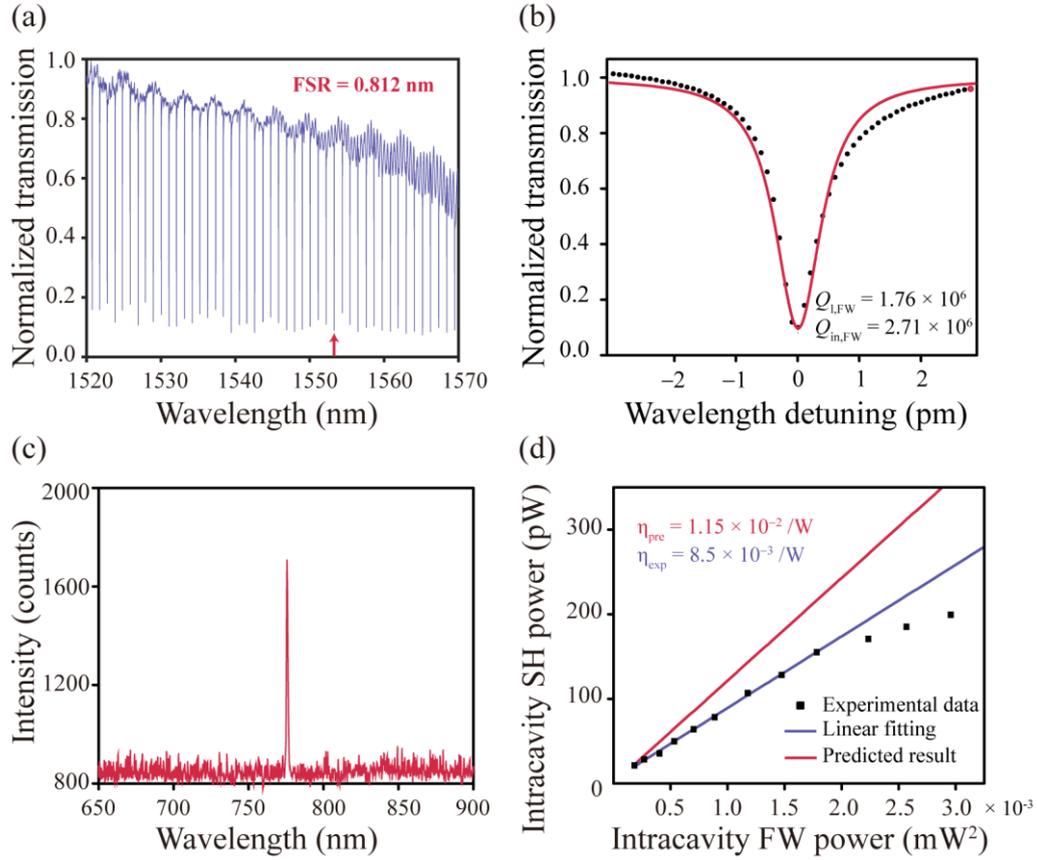

**Fig. 4 Experimental demonstration of the 111st-order SQPM SHG.** (a) Transmission spectrum for the $TE_{00}$ mode of FW. The red arrow marks the mode used in the experiment on SHG. (b) Lorentz fitting for the marked mode with a centre wavelength of 1553.01 nm. (c) Spectrum of the SH signal. (d) Measured data, linear fitting, and theoretical prediction of the SHG intensity versus the square of intracavity FW power.

Fig. 4(a) shows the transmission spectrum for the FW. The free spectral range (FSR) near 1550 nm is about 0.812 nm. The marked dip is fitted with the Lorentz function in Fig. 4(b). Based on this, the loaded and intrinsic Q factors were calculated as $1.76 \times 10^6$ and $2.71 \times 10^6$, respectively, implying a propagation loss of 0.12 dB/cm at 1550 nm band approximately. The spectrum for the SH is shown in Fig. 4(c), and the variation in the intracavity power for the SH and FW is plotted in Fig. 4(d). The extracted conversion efficiency of SHG is about $8.5 \times 10^{-3}$/W. As the FW power increases, the measured data begin to deviate from linearity, and we attribute this phenomenon to thermal detuning of the FW. The ideal conversion efficiency for our experiment is predicted by Eqs. (5) and (6), with the assumption that $Q_{l,SH}$ ($\sim 2 \times 10^5$) is lower than $Q_{l,FW}$ but the difference is within one order of magnitude, which has been confirmed in previous works[31, 32]. The red line in Fig. 4(d) shows the theoretically predicted result ($\sim 1.15 \times 10^{-2}/W$). It was found that the experimental conversion efficiency was nearly 80 percent of the predicted value; this is mainly a result of fabrication errors, which causes imperfect SQPM and a conservative estimate of the intracavity power of SH.

# Discussion

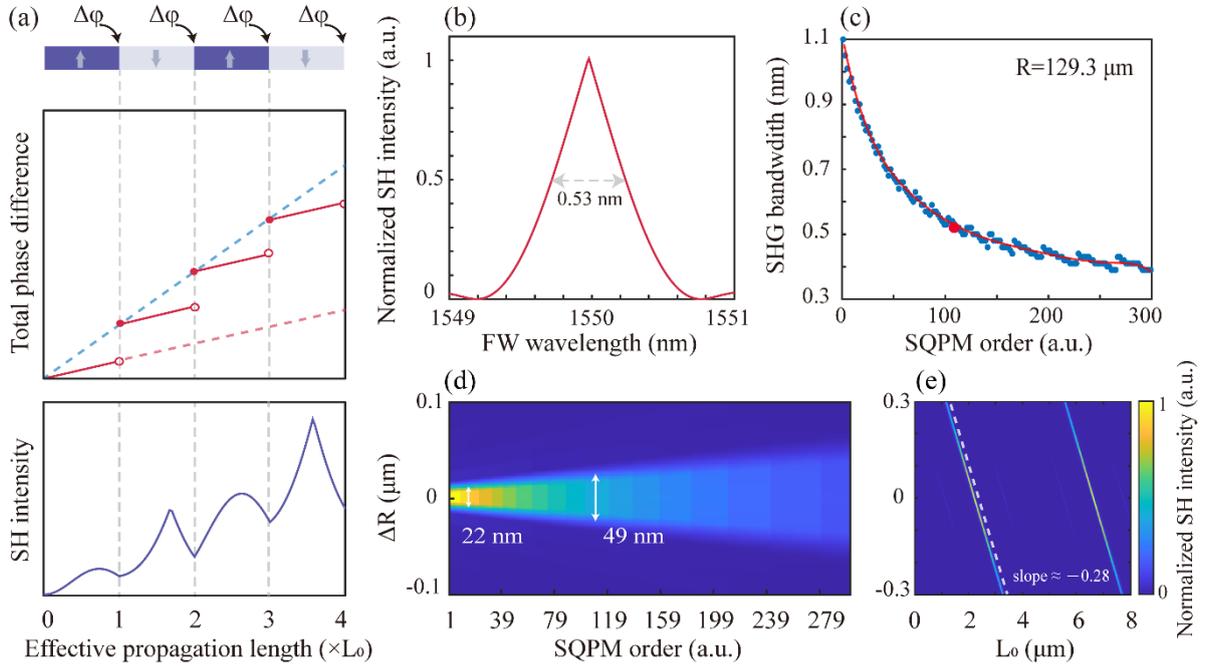

**Fig. 5 Imperfect SQPM SHG in the micro-racetrack resonator.** (a) Diagram of the phase difference between the FW and SH and corresponding variation of the SH intensity when the operating state of SQPM micro-racetrack resonator derivates a little from the perfect phase-matching. (b) Numerical simulated SH intensity spectrum in our designed SQPM structure with the wavelength of FW varying from $1550\ nm$ to $1551\ nm$. (c) Calculated bandwidths versus different order-numbers of SQPM with $R = 129.3\ \mu m$, where the red dot represents a theoretical estimation of the bandwidth of $0.52\ nm$ in this work. (d)Simulated SH intensity distribution at $775\ \mu m$ with the deviation of $R$ from $129.3\ \mu m$ and different SQPM order, (e)or continuously varied straight waveguide length $L_0$.

In practice, varies factors like the deviation in the FW wavelength and the fabrication errors will perturb the ideal SQPM condition, bringing negative effects in two aspects: (1) incomplete compensation of the phase-mismatching in the straight waveguide sections and (2) additional phase delay introduced by the half-ring waveguides. Here, we can treat the SQPM micro-racetrack resonator as a periodical poled straight radge waveguide with a poling period of $2L_0$ and an equivalent length of $L_{eqv} = 2L_0N$, where $N$ is an integer representing the maximum number of the circle the light can circulate in the racetrack resonator. It is noteworthy that the path in the half-ring waveguide is not considered when we calculate the effective propagation length that contributes to the SHG process. As the diagram in the Fig. 5(a) has shown, a phase delay of $\Delta\varphi$, which no longer equals to the integra multiplies of $2\pi$, adds to the total phase difference between the FW and SH at the boundary of equivalent domains. Consequently, the total phase difference under an imperfect SQPM condition is illustrated in the middle graph of Fig. 5 (a). The phase difference at the start points of each equivalent domain is determined by the blue dashed line which has a slope of $\Delta k' = \Delta k + \Delta\varphi/L_0$, while in each equivalent domain, it grows linearly with a rate of $\Delta k$ as the effective propagation length increases. Hence, based on the phase information and the coupled-mode equation, SH intensity in the SQPM racetrack resonator can be calculated numerically. The bottom graph in Fig. 5(a) shows a schematic of the intensity variation of SH under the imperfect SQPM condition, where the SH intensity goes through a rather complex process instead of continuous growth.

In Fig. 5(b), we calculate the SH intensity spectrum when the wavelength of input FW is tuned from $1549\ nm$ to $1551\ nm$ with the designed racetrack structure ($R = 129.3\ \mu m$, $L_0 = 111\ L_c$). The SHG bandwidth is approximately $0.53\ nm$ in this work. Then we change the SQPM order from the 1st to the 301st order while keep $R$ unchanged, corresponding bandwidths is shown in Fig. 5(c). As the SQPM order gets lower, the bandwidth is broadened obviously. And for the 1st order SQPM, the bandwidth reaches $1.1\ nm$ approximately. Fig. 5(d) shows the SH intensity at the FW wavelength of $1550\ nm$ with the $R$ deviated from designed $129.3\ \mu m$ and $L_0$ varies as the discrete values to meet the requirement of different-order SQPM. As we can see, the tolerance of $R$ is approximately $49\ nm$ in this work (111st order SQPM), and it is much larger than the fabrication error (about $10\ nm$) associated with the waveguide width as well as the outside radius $R$ of the half-ring waveguide. Even for a shorter $L_0$ with an SQPM order number lower than 15, the tolerance is about $22\ nm$. At last, we investigate the SH intensity with both the $R$ and $L_0$ deviated from the designed values, as shown in Fig. 5(e). Here $L_0$ continuously varies from $0\ \mu m$ to $8\ \mu m$, covering the 1st and 3rd order SQPM. It is interesting to find that the shape of the bright regime, which can be regarded as a symbol of phase-matching, is a line instead of a rectangle, indicating an underlying kind of phase-matching in our SQPM racetrack resonator, in which the phase-mismatching in the straight waveguide section just can be compensated by the additional phase delay $\Delta\varphi$, e.g., $\Delta k L_0 + \Delta\varphi = m\pi$, thus the intensity of SH will grow continuously, but less rapidly than the completely fulfilled SQPM condition. Since $R$ varies in a range smaller than $1\ \mu m$, we can assume that the effective refractive index in the half-ring waveguide is constant, thus $\Delta\varphi$ will directly depend on $R$ by $\Delta\varphi = \frac{2\pi}{\lambda}\int_0^\pi n_{eff,FW}\, R d\theta$, and a linear expression associated with $R$ and $L_0$ under such underlying phase-matching condition can be easily derived, which agrees well with the simulated result in Fig. 5(e).

In above discussion of the practical SQPM enabled SHG process, we ignore the resonance condition in the resonator for simplicity. However, limited by the fabrication accuracy, the designed FW wavelength for SQPM SHG may deviate from the resonant mode of the micro-racetrack resonator, as shown in Fig. 6(a). Fortunately, if the FSR is close to or smaller than the SHG bandwidth, the resonant modes are sufficiently dense, and at least one resonant mode can be located within the effective range of SHG. By means of thermo-optic or electro-optic modulation, the resonant mode can be made to perfectly match the central wavelength. In practice, as we have discussed in the last section, the SHG bandwidth will be enlarged for a lower SQPM order. By adopting a shorter $L_0$ and designing the $R$ in a proper size, it is possible to make the SHG bandwidth comparable with, even larger than the FSR.

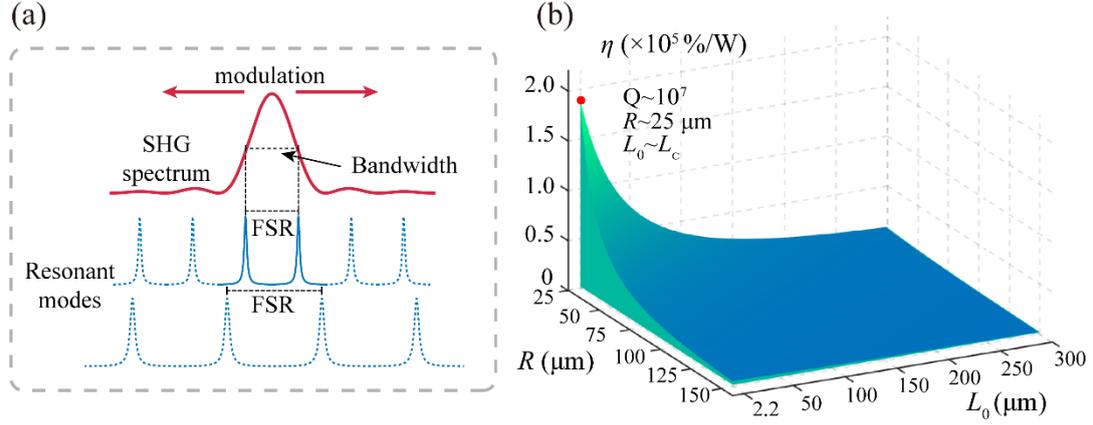

**Fig. 6 Further improvement of the SQPM micro-racetrack resonator.** (a) Dependence of the matching and mismatching conditions between the resonant modes and the SHG central wavelength on the FSR. The central wavelength of the SHG spectrum can shift under thermo-optic or electro-optic modulation of the straight waveguides. (b) Predicted normalized intracavity conversion efficiency of SHG in the SQPM micro-racetrack resonator for different half-ring waveguide radii (25 ~ 150 μm) and straight waveguide lengths ($L_c$ ~ $135L_c$) with a loaded Q factor of $10^7$.

At last, we try to explore the performance ceiling of SQPM micro-racetrack resonator in the nonlinear frequency conversion process. According to Eq. (3), normalized intracavity conversion efficiency can be approximated as

$$\eta_0 \propto \frac{Q_{l,SH}^2}{(\pi R + L_0)^2}, \tag{7}$$

which indicates that micro-racetrack resonators with smaller R and $L_0$ can provide a higher SHG efficiency. For a loaded Q factor maintained at a moderate value of $10^7$, we calculate the SHG efficiency in the SQPM micro-racetrack resonator with different geometrical parameters, as shown in Fig. 6(b). The red dot indicates that the highest intracavity normalized conversion efficiency of SQPM-enabled SHG is predicted to be 186,000%/W for a minimum radius of 25 μm and a minimum straight waveguide length of $L_c$, corresponding to the 1st order SQPM. Although this value can be influence by the phase-matching and resonant conditions, it still reveals huge potential of SQPM in the nonlinear frequency conversion on integrated photonics platform. With an optimized waveguide-resonator coupling efficiency up to 0.25, an on-chip conversion efficiency of approximately 2906%/W is predicted for application in practice.

## Materials and Methods

### Simulation

The effective refractive indices in the straight and half-circle waveguides are calculated using the finite element method. To calculate the phase delay in the half-circle waveguide, we first divide the half-ring waveguide into 60 tiny curved waveguides, where the material refractive index of the TE-polarized light in each of them is

$$n_e(\theta) = \left(\frac{\cos^2\theta}{n_o^2} + \frac{\sin^2\theta}{n_e^2}\right)^{-1/2}, \tag{8}$$

where $\theta$ is the azimuth angle defined in Fig. 1, and $n_o$ and $n_e$ are the ordinary and extraordinary refractive indices of lithium niobate, respectively. We then use the conformal transformation[38] to transfer the curved waveguide into an equivalent straight waveguide and conduct a mode analysis simulation. In this way, the phase delay of FW and SH in each tiny curved waveguide can be calculated exactly, and the total can be easily obtained by summation.

### Device fabrication

The micro-racetrack resonator was fabricated on a 600 nm thick thin film lithium niobate (by NanoLN). An amorphous silicon thin film was deposited on the thin film lithium niobate as a hard mask via plasma-enhanced chemical vapour deposition, and the designed patterns were defined in ZEP520A resist using standard electron beam lithography. We first transferred the pattern to the hard mask by reactive ion etching with sulfur hexafluoride, and subsequently to the lithium niobate film by inductively coupled plasma-reactive ion etching with argon. Finally, RCA cleaning was performed to remove the hard mask and small particles.

### Intracavity power estimation

In the experiment, the SH signal was directly detected by the OSA in the form of the photon number N, and the input FW power $P_{\text{input}}$ was measured by a power meter before being coupled to the grating coupler. The intracavity power for FW and SH was estimated as

$$P_{\text{in,FW}} = P_{\text{input}} g_{\text{FW}} \kappa_{\text{FW}} \alpha_{\text{FW}} \tag{9}$$

and

$$P_{\text{in,SH}} = \frac{Nh\nu}{\Delta t\, g_{\text{SH}} \kappa_{\text{SH}} \alpha_{\text{SH}}}, \tag{10}$$

where $g_{FW(SH)}$, $\kappa_{FW(SH)}$, and $\alpha_{FW(SH)}$ are the grating coupling efficiency, the waveguide-resonator coupling efficiency, and the attenuation factor in the optical path of FW (SH), respectively. $h\nu$ is the single photon energy for SH, and $\Delta t$ is the integration time of optical spectrum analyser, which was set to $0.5\ s$ in our experiment. The grating coupling efficiencies of FW and SH were measured as $g_{FW} \sim 0.2$ and $g_{SH} \sim 0.016$. The waveguide-resonator coupling efficiency for FW was calculated as $\kappa_{FW} \sim 0.0027$ based on the coupling Q-factor measured in the experiment, and $\kappa_{SH}$ was also assumed to be 0.0027. In practice, $\kappa_{FW}$ should be much higher than $\kappa_{SH}$, since the coupling structure was specifically designed to couple the FW fundamental mode between the waveguide and resonator. However, we made this assumption to avoid overestimating the intracavity power of SH, and the real conversion efficiency in the cavity is likely to be higher. Finally, the attenuation factors for FW and SH were estimated as 0.8, as an empirical value.

## Conclusion

In this paper, we have theoretically and experimentally demonstrated an SQPM SHG in a fully integrated micro-racetrack resonator based on dispersion engineering and the design of the crystal orientation. The intracavity conversion efficiency could be improved by optimizing the racetrack structure, improving the resonator Q-factor, or applying thermo-optic or electro-optic modulation to the resonance wavelength. By using the first-order QPM, our theoretical analysis predicted that the intracavity conversion efficiency could reach $186,000\%/W$, which can be applied in practice with an optimized design of outward coupling. In addition, the SHG bandwidth and fabrication tolerance of SQPM micro-racetrack resonator is also investigated. This novel approach to achieve a chip-scale efficient nonlinear frequency conversion process is compatible with standard CMOS technology, and is suitable for large-scale production of integrated photonic chips. By designing the racetrack structures to compensate for different kinds of phase mismatching, SQPM can be employed for almost all second-order nonlinear optical processes, which will be valuable in terms of fabricating integrated quantum light sources or other nonlinear photonic elements on integrated photonic circuit.

# References

[1] Christian Kurtsiefer, Markus Oberparleiter, and Harald Weinfurter. High efficiency entangled photon pair collection in type II parametric fluorescence. *Physical Review A* **64**(2), 023802 (2001).

[2] Rui Luo, Haowei Jiang, Steven D. Rogers, Hanxiao Liang, Yang He, and Qiang Lin. On-chip second-harmonic generation and broadband parametric down-conversion in a lithium niobate microresonator. *Optics Express* **25**(20), 24531–24539 (2017).

[3] Zhaohui Ma, Jia yang Chen, Zhan Li, Chao Tang, Yong Meng Sua, Heng Fan, and Yu-Ping Huang. Ultrabright quantum photon sources on chip. *Physical Review Letters* **125**(26), 263602 (2020).

[4] Ling-An Wu, H. J. Kimble, J. L. Hall, and Huifa Wu. Generation of squeezed states by parametric down conversion. *Phys. Rev. Lett.* **57**(20), 2520–2523 (1986).

[5] Xiang Guo, Chang ling Zou, Carsten Schuck, Hojoong Jung, Risheng Cheng, and Hong X. Tang. Parametric down-conversion photon-pair source on a nanophotonic chip. *Light, Science & Applications* **6**(5), e16249-e16249 (2017).

[6] Kumar, P. Quantum frequency conversion. *Optics Letters* **15**(24), 1476–8 (1990).

[7] Michael G. Raymer and Kartik Srinivasan. Manipulating the color and shape of single photons. *Physics Today* **65**(11), 32–37 (2012).

[8] Robert W Boyd, R. W. Nonlinear Optics (Academic Press, 2020).

[9] JA Armstrong, N Bloembergen, J Ducuing, and Peter S Pershan. Interactions between light waves in a nonlinear dielectric. *Physical Review*, **127**(6), 1918 (1962).

[10] P. A. Franken, and J. F. Ward. Optical harmonics and nonlinear phenomena. *Rev. Mod. Phys.* **35**, 23–39 (1963).

[11] Shi Ning Zhu, Yong-yuan Zhu, and Nai-ben Ming. Quasi-phase-matched third-harmonic generation in a quasi-periodic optical superlattice. *Science* **278**(5339), 843-846(1997).

[12] M.M. Fejer, G.A. Magel, D.H. Jundt, and R.L. Byer. Quasi-phase-matched second harmonic generation: Tuning and tolerances. *IEEE Journal of Quantum Electronics* **28**(11), 2631–2654 (1992).

[13] Sina Saravi, Thomas Pertsch, and Frank Setzpfandt. Lithium niobate on insulator: An emerging platform for integrated quantum photonics. *Advanced Optical Materials* **9**(22), 2100789 (2021).

[14] Cheng Wang, Mian Zhang, Xi Chen, Maxime Bertrand, Amirhassan Shams-Ansari, Sethumadhavan Chandrasekhar, Peter Winzer, and Marko Lončar. Integrated lithium niobate electro-optic modulators operating at CMOS-compatible voltages. *Nature* **562**(7725), 101–104 (2018).

[15] Mingbo He, Mengyue Xu, Yuxuan Ren, Jian Jian, Ziliang Ruan, Yongsheng Xu, Shengqian Gao, Shihao Sun, Xueqin Wen, Lidan Zhou, et al. High-performance hybrid silicon and lithium niobate Mach–Zehnder modulators for 100 Gbit s$^{-1}$ and beyond. *Nature Photonics* **13**(5), 359–364 (2019).

[16] Mingxiao Li, Jingwei Ling, Yang He, Usman A Javid, Shixin Xue, and Qiang Lin. Lithium niobate photonic-crystal electro-optic modulator. *Nature Communications* **11**(1), 1–8 (2020).

[17] Linbo Shao, Mengjie Yu, Smarak Maity, Neil Sinclair, Lu Zheng, Cleaven Chia, Amirhassan Shams-Ansari, Cheng Wang, Mian Zhang, Keji Lai, et al. Microwave-to-optical conversion using lithium niobate thin-film acoustic resonators. *Optica* **6**(12), 1498–1505 (2019).

[18] YiAn Liu, XiongShuo Yan, JiangWei Wu, Bing Zhu, YuPing Chen, and XianFeng Chen. On-chip erbium-doped lithium niobate microcavity laser. *Science China: Physics, Mechanics & Astronomy* **64**(3), 1–5 (2021).

[19] Zhe Wang, Zhiwei Fang, Zhaoxiang Liu, Wei Chu, Yuan Zhou, Jianhao Zhang, Rongbo Wu, Min Wang, Tao Lu, and Ya Cheng. On-chip tunable microdisk laser fabricated on Er$^{3+}$-doped lithium niobate on insulator. *Optics Letters* **46**(2), 380–383 (2021).

[20] Qiang Luo, ZhenZhong Hao, Chen Yang, Ru Zhang, DaHuai Zheng, ShiGuo Liu, HongDe Liu, Fang Bo, YongFa Kong, GuoQuan Zhang, et al. Microdisk lasers on an erbium-doped lithium-niobite chip. *Science China: Physics, Mechanics & Astronomy* **64**(3), 1–5 (2021).

[21] DiFeng Yin, Yuan Zhou, Zhaoxiang Liu, Zhe Wang, Haisu Zhang, Zhiwei Fang, Wei Chu, Rongbo Wu, Jianhao Zhang, Wei Chen, et al. Electro-optically tunable microring laser monolithically integrated on lithium niobate on insulator. *Optics Letters* **46**(9), 2127–2130 (2021).

[22] Zhaoxi Chen, Qing Xu, Ke Zhang, WingHan Wong, DeLong Zhang, Edwin YueBun Pun, and Cheng Wang. Efficient erbium-doped thin-film lithium niobate waveguide amplifiers. *Optics Letters* **46**(5), 1161–1164 (2021).

[23] Minglu Cai, Kan Wu, Junmin Xiang, Zeyu Xiao, Tieying Li, Chao Li, and Jianping Chen. Erbium-doped lithium niobate thin film waveguide amplifier with 16 dB internal net gain. *IEEE Journal of Selected Topics in Quantum Electronics* **28**(3), 1-8 (2021).

[24] Zhe Wang, Chaohua Wu, Zhiwei Fang, Min Wang, Jintian Lin, Rongbo Wu, Jianhao Zhang, Jianping Yu, Miao Wu, Wei Chu, Tao Lu, Gang Chen, and Ya Cheng, High-quality-factor optical microresonators fabricated on lithium niobate thin film with an electro-optical tuning range spanning over one free spectral range, *Chinese Optics Letters* **19**(6), 060002(2021)

[25] Ke Zhang, Zhaoxi Chen, Hanke Feng, Wing-Han Wong, Edwin Yue-Bun Pun, and Cheng Wang, High-Q lithium niobate microring resonators using lift-off metallic masks, *Chinese Optics Letters* **19**(6), 060010 (2021)

[26] Vahala, Kerry J. Optical microcavities, *nature* **424**(6950), 839-846(2003)

[27] Lin, Guoping, Aurélien Coillet, and Yanne K. Chembo. Nonlinear photonics with high-Q whispering-gallery-mode resonators, *Advances in Optics and Photonics* **9**(4), 828-890(2017)

[28] Xueyue Zhang, Qi-Tao Cao, Zhuo Wang, Yu-xi Liu, Cheng-Wei Qiu, Lan Yang, Qihuang Gong and Yun-Feng Xiao, Symmetry-breaking-induced nonlinear optics at a microcavity surface, *Nature Photonics* **13**(1), 21-24(2019)

[29] Jin-hui Chen, Xiaoqin Shen, Shui-Jing Tang, Qi-Tao Cao, Qihuang Gong, and Yun-Feng Xiao, Microcavity nonlinear optics with an organically functionalized surface, *Physical Review Letters* **123**(17), 173902(2019)

[30] Rui Ge, Xiongshuo Yan, Yuping Chen, and Xianfeng Chen, Broadband and lossless lithium niobate valley photonic crystal waveguide, *Chinese Optics Letters* **19**(6), 060014(2021)

[31] JiaYang Chen, ZhaoHui Ma, Yong Meng Sua, Zhan Li, Chao Tang, and YuPing Huang. Ultra-efficient frequency conversion in quasi-phase-matched lithium niobate microrings. *Optica* **6**(9), 1244–1245 (2019).

[32] Juanjuan Lu, Joshua B Surya, Xianwen Liu, Alexander W Bruch, Zheng Gong, Yuntao Xu, and Hong X Tang. Periodically poled thin-film lithium niobate microring resonators with a second-harmonic generation efficiency of 250,000%/W. *Optica* **6**(12), 1455–1460 (2019).

[33] Juanjuan Lu, Ming Li, Chang-Ling Zou, Ayed Al Sayem, and Hong X. Tang. Toward 1% single-photon anharmonicity with periodically poled lithium niobate microring resonators. *Optica* **7**(12), 1654–1659 (2020).\

[34] Dehui Sun, Yunwu Zhang, Dongzhou Wang, Wei Song, Xiaoyan Liu, Jinbo Pang, Deqiang Geng, Yuanhua Sang and Hong Liu, Microstructure and domain engineering of lithium niobate crystal films for integrated photonic applications. *Light: Science & Applications*, **9**(1), 1-18 (2020).

[35] Ehsan Shah Hosseini, Siva Yegnanarayanan, Amir Hossein Atabaki, Mohammad Soltani, and Ali Adibi. Systematic design and fabrication of high-Q single-mode pulley-coupled planar silicon nitride microdisk resonators at visible wavelengths. *Optics Express* **18**(3), 2127–2136 (2010).

[36] Amnon Yariv. Universal relations for coupling of optical power between microresonators and dielectric waveguides. *Electronics Letters* **36**, 321–322 (2000).

[37] Boris Sturman and Ingo Breunig. Generic description of second-order nonlinear phenomena in whispering-gallery resonators. *Journal of the Optical Society of America B: Optical Physics* **28**, 2465–2471 (2011).

[38] Mordehai Heiblum, M. & Harris, J. Analysis of curved optical waveguides by conformal transformation. *IEEE Journal of Quantum Electronics* **11**(2), 75–83 (1975).

## Data availability

The data that support the findings of this study are available from the corresponding author upon reasonable request.

## Acknowledgements

This work was supported by the National Key R&D Program of China (Grant Nos. 2019YFB2203501 and 2017YFA0303701), the National Natural Science Foundation of China (Grant Nos. 12134009, and 91950107), Shanghai Municipal Science and Technology Major Project (2019SHZDZX01-ZX06), and SJTU No. 21X010200828.

## Author contributions

Q.L. and H.W.J. conceived the idea. Y.P.C. and X.F.C. supervised the research. Y.A.L. performed the fabrication and SHG experiment. X.S.Y. and R.G. helped with the theoretical analysis and experiment. H.L. helped with the fabrication of the device. T.G.Y. performed the design, simulation, theoretical analysis and prepared the manuscript with the assistance of all the co-authors.

## Competing interests

The authors declare no competing interests.

# Supplemental Material

This document provides the derivation of the intracavity conversion efficiency of SQPM enabled SHG process (Eq. (5) and (6) in the main text). The time-dependent variation of the SH in the cavity can be described as

$$\frac{dA_{SH}}{dt} = -\frac{1}{2}\left(\frac{1}{\tau_{in}} + \frac{1}{\tau_c}\right)A_{SH} + \frac{\Delta A_{SH}}{T_{SH}} \quad (1)$$

where $A_{SH}$ is the intracavity amplitudes of SH, and its evolution is mainly dominated by three factors: the intrinsic optical loss of the cavity, coupling loss with the pulley waveguide and the gain from the nonlinear frequency conversion process [1]. We use the intrinsic photon lifetime $\tau_{in}$ and the coupling photon lifetime $\tau_c$ to characterize the decay behavior of the SH in the cavity, and use $\frac{\Delta A_{SH}}{T_{SH}}$ to denote the nonlinear effect induced amplitude growth rate.

The theoretical intracavity conversion efficiency is derived under the steady states of SH. When $\frac{dA_{SH}}{dt} = 0$, the intracavity amplitude of SH can be calculated as

$$A_{SH} = 2\Delta A_{SH} / \left[\left(\frac{1}{\tau_{in}} + \frac{1}{\tau_c}\right)T_{SH}\right], \quad (2)$$

where $\Delta A_{SH}$ is an intracavity FW amplitude-dependent term, which is calculated from the nonlinear coupled wave equation of SHG in the undepleted regime [2] as

$$\Delta A_{SH} = \frac{8 d_{33} L_c}{\lambda_{FW} n_{eff,SH}} A_{FW}^2, \quad (3)$$

where $A_{FW}$ is the intracavity amplitude of the FW; $T_{SH}$ is the time period of SH traveling in the micro-racetrack resonator, which can be calculated by $T_{SH} = (m_{azi,SH} + M_{SH})\frac{\lambda_{SH}}{c}$. In experiment, $\tau_{in}$ and $\tau_c$ are characterized by the intrinsic and coupling Q-factor with the equation of $\tau_{in(c)} = \lambda_{SH} Q_{in(c)}/2\pi c$. Since $Q_l^{-1} = Q_{in}^{-1} + Q_c^{-1}$, $\frac{1}{\tau_{in}} + \frac{1}{\tau_c}$ can be expressed by the loaded Q-factor $Q_{l,SH}$ as $2\pi c/\lambda_{SH} Q_{l,SH}$. Finally, taking the square of both sides in Eq. (1), we have

$$A_{SH}^2 = \frac{64 d_{33}^2 L_c^2 Q_{l,SH}^2}{\pi^2 \lambda_{FW}^2 n_{eff,SH}^2 \left(m_{azi_{SH}} + M_{SH}\right)^2} A_{FW}^4. \quad (4)$$

The relation between the amplitude and the power is described by $P = \frac{1}{2}\epsilon_0 c n \iint_\sigma |A|^2 d\sigma$, where $\sigma$ is the cross-section area of the mode in the straight waveguide. Thus, we can calculate the intracavity power of SH and FW by

$$P_{SH} = \frac{1}{2}\epsilon_0 c n_{eff,SH} \iint A_{SH}^2 \, d\sigma_{SH} \quad (5)$$

and

$$P_{FW} = \frac{1}{2}\epsilon_0 c n_{eff,FW} \iint A_{FW}^2 \, d\sigma_{FW}, \quad (6)$$

respectively. Combining Eq. (4-6), the intracavity conversion efficiency is derived as

$$\eta = \frac{P_{SH}}{P_{FW}^2} = \eta_0 \frac{\iint |A_{FW}|^4 d\sigma_{SH}}{[\iint |A_{FW}|^2 d\sigma_{FW}]^2} \quad (7)$$

with

$$\eta_0 = \frac{128 d_{33}^2 L_c^2 Q_{l,SH}^2}{c\epsilon_0 \pi^2 \lambda_{FW}^2 n_{eff,FW}^2 n_{eff,SH} \left(m_{azi_{SH}} + M_{SH}\right)^2}. \tag{8}$$

As a result, once the structure information of the micro-racetrack resonator and the loaded Q-factor at SH band are known, the intracavity conversion efficiency of the SHG process in the SQPM micro-racetrack resonator can be predicted.

**Reference**

[1] Boris Sturman and Ingo Breunig. Generic description of second-order nonlinear phenomena in whispering gallery resonators. Journal of The Optical Society of America B-optical Physics, 28:2465–2471, 2011.

[2] Robert W Boyd. Nonlinear optics. Academic press, 2020.